\documentclass[11pt]{article}
\usepackage{amsmath}
\usepackage{graphicx}
\usepackage{amssymb}
\usepackage{amsfonts}
\usepackage{latexsym}

\def\beq{\begin{eqnarray}}
\def\eeq{\end{eqnarray}}

\def\al{\alpha}
\def\be{\beta}

\def\ga{\gamma}\def\de{\delta}

\def\ep{\epsilon}

\def\ka{\kappa}
\def\la{\lambda}

\def\si{\sigma}

\def\Si{\Sigma}

\begin{document}

\title{On the action for Weyssenhoff spin fluid and the
Barbero-Immirzi parameter}
\author{Guilherme de Berredo-Peixoto\footnote{Email address:
guilherme@fisica.ufjf.br}
$\,$ and $\,$ Cleber A. de Souza\footnote{Email address:
abrahaocleber@gmail.com }\\ \\
Departamento de F\'{\i}sica, ICE, Universidade Federal de Juiz de Fora
\\
Campus Universit\'{a}rio - Juiz de Fora, MG, Brazil  36036-330}
\date{}
\maketitle

\begin{quotation}
\noindent
{\large{\bf Abstract.}}

 It was showed by Perez and Rovelli in 2006 that the Holst action
 in gravity with torsion with massless and minimally coupling Dirac
 fermions gives rise to the four-fermion coupling term, whose coefficient
 is a function of the Barbero-Immirzi (BI) parameter. This parameter
 is present in the Holst action, which is an object of investigation
 in a non-perturbative formalism of quantum gravity. The key feature
 is the torsion because its absence implies no effect from Holst term
 in dynamical equations. In this paper we consider a spin fluid (also
 called Weyssenhoff fluid), which is a perfect fluid that has intrinsic
 spin. We study the Host action in gravity with torsion and spin fluid,
 and we include, for completeness, minimaly coupling massive fermions.
 We find the equivalent action containing the same four-fermion interaction
 term previously calculated by Perez and Rovelli without the spin fluid,
 also the quadratic spin tensor term and finally
 an interaction term between the axial current and the spin tensor,
 which can be relevant for new effects and features in Cosmology. In
 all three cases, the dependence on BI parameter is the same.

\vskip 4mm
PACS: $\,$
04.20.-q    
$\,\,$
98.80.-k    
$\,\,$
04.50.Kd    
\vskip 1mm

Keywords:  Dirac fields, Torsion, Barbero-Immirzi parameter, Weyssenhoff Fluid

\end{quotation}

\section{Introduction}

Dirac fields in curved spacetime is a subject which has been
studied in a number of works for many years, especially
in the last decade, as shown in Refs.
\cite{deser,epstein,kremer2,cheng,boeckel,
chimentokremer,samojeden,rakhi,rakhi2,fabbri}.
In this paper, we consider classical aspects
of Dirac particles in curved spacetime in the
presence of the BI parameter \cite{barbero,immirzi} and the Weyssenhoff fluid.
The BI parameter, $\be$, carried by the Holst action term \cite{holst},
$$
\frac{1}{2\be} \,
\ep_{\al\be\mu\nu} \,R^{\al\be\mu\nu}\,,
$$
where $\ep_{\al\be\mu\nu}$ is the Levi-Civita
tensor, was introduced in the non-perturbative quantum gravity scenario,
and is indeed a new dimensionless parameter emerging from
a more fundamental theory. In General Relativity (GR) this parameter
has no dynamical effect, because $\ep_{\al\be\mu\nu} R^{\al\be\mu\nu} = 0$
by ciclic symmetry of the Riemann tensor
($R^{\al\be\mu\nu} + R^{\al\nu\be\mu} + R^{\al\mu\nu\be} = 0$). The situation
changes for the non-Riemannian case of Einstein-Cartan action, where
$\tilde{R}^{\al\be\mu\nu} + \tilde{R}^{\al\nu\be\mu} +
\tilde{R}^{\al\mu\nu\be} \neq 0$. Here we represent curvatures with tilde
as the curvature constructed with torsion. For an introduction and review on
the theories with torsion, see for example references
\cite{revhehl,torsi,Puetzfeld}. In the theory with Einstein-Hilbert and Holst
action, together with minimal coupling fermions in the presence of torsion,
the BI parameter does affect the gravitational dynamics, providing an
interesting way to investigate its classical and quantum effects, through
4-fermion interaction term \cite{perezrovelli,BPFSS}. In Refs.
\cite{freidel,mercuri,alexandrov,taveras,lagraa} another coupling scenarios
and features are explored on the issue of the role of BI parameter in physical
theories.

The Weyssenhoff fluid \cite{weyssenhoff} is a fluid with intrinsic
spin density, which is described by the spin tensor $\Sigma_{\al\be}$.
In principle, this fluid can not be associated with a Lagrangian description
based on spinor fields. One has to postulate spin extra terms in the
energy-momentum tensor. There are several papers considering this
fluid, most of all concerned with cosmological applications as a
realization of torsion effects in Cosmology (see, for example,
\cite{emanuel} and references therein). One can cite, for instance,
the work of Gasperini \cite{gasperini} which considered the
energy momentum tensor (of the Weyssenhoff fluid) previoulsy improved by
Ray and Smalley \cite{ray}, where spin is a thermodynamical variable.
In Ref. \cite{gasperini}, the torsion, algebraically related to the
quantities describing the sources (what happens in most of the papers
in this theme, including this one), provides
singularity avoidance and early accelerated expansion,
but the expansion factor of the cosmological scale is too small.
Obukhov and Korotky \cite{obukhov} formulated
a more general variational theory describing the Weyssenhoff fluid and
also applied to cosmological models with rotation, shear and
expansion.

In this paper we study the effect of the Holst action in a spacetime
with torsion together with minimally coupling Dirac fields and also
a Weyssenhoff fluid. We find the equivalent action which is composed
by the Riemannian Einstein-Hilbert term and the following source terms:
a four-fermion interaction expression, $J^\mu J_\mu $, a quadratic
spin tensor term, $\Sigma_{\mu\nu}\Sigma^{\mu\nu}$ and a mixing term
(which can provide an interesting study of new effects in Cosmology).
The coefficients of these three terms are proportional to the same
function of the BI parameter. In the next section, we write the
starting action and some basic definitions, then we derive the
equations with respect to torsion in Section 3. The conclusions is
then drawn in Section 4.

\section{Holst action with Dirac fields and Weyssenhoff fluid}

Our starting point is the Holst action in the presence of torsion,
as well as minimally coupling fermions and the Weyssenhoff fluid:

\beq
S = S_H + S_\Sigma + S_\psi + S_{PF}\,,
\label{action}
\eeq
where $S_H$ is the action for Einstein-Cartan theory with also the Holst
term, $S_\Sigma$ is the action describing the interaction between torsion
and the intrinsic spin of the fluid, $\Sigma_{\mu\nu}$, the action $S_\psi$
is the minimally coupled Dirac action and $S_{PF}$ is the action for the
perfect fluid counterpart of the Weyssenhoff fluid such that the
corresponding energy-momentum tensor is the standard one, $T_{\mu\nu} =
(p+\rho)U_\mu U_\nu - p\, g_{\mu\nu}$.

The first contribution to the total action is the one for gravitational sector, $S_H$:
\beq
S_H = \frac{1}{\ka}\int d^4x\sqrt{-g}\left\{ -\tilde{R} +
\frac{1}{2\be}\ep^{\mu\nu\al\be}\tilde{R}_{\mu\nu\al\be}\right\}\,.
\eeq
Here $\ka = 16\pi G$, $\ep^{\mu\nu\al\be}$ is the Levi-Civita tensor and
the tilde above curvatures means that they are constructed using the
full connection $\tilde{\Gamma}^\rho\mbox{}_{\al\be}$ which includes the
torsion tensor
$$
T^\rho\mbox{}_{\al\be} = \tilde{\Gamma}^\rho\mbox{}_{\al\be} -
\tilde{\Gamma}^\rho\mbox{}_{\be\al}\,.
$$
Using the above definition, the metricity condition gives us the expression
of the full connection in terms of the Riemannian connection,
$\Gamma^\rho\mbox{}_{\al\be}$, as
$$
\tilde{\Gamma}^\rho\mbox{}_{\al\be} = \Gamma^\rho\mbox{}_{\al\be} + K^\rho\mbox{}_{\al\be}\,,
$$
where the last term is called contortion tensor, given by
$$
K^\rho\mbox{}_{\al\be} = \frac{1}{2}\left(T^\rho\mbox{}_{\al\be} -
T_\al\mbox{}^\rho\mbox{}_\be - T_\be\mbox{}^\rho\mbox{}_\al\right)\,.
$$

The second term in the right hand side of equation (\ref{action}) is the
action for the interaction between torsion and spin of the fluid \cite{ray}:
\beq
S_\Sigma = \al\int d^4x\sqrt{-g}\Sigma^{\mu\nu}\,U^\la\,K_{\mu\nu\la}\,,
\eeq
where $U^\mu$ is the 4-velocity in comoving frame, $U^\mu = (1,0,0,0)$ and
$\Sigma^{\mu\nu}$ is the antisymmetric spin tensor of the fluid which satisfies
the Frenkel condition $\Sigma^{\mu\nu}U_\nu = 0$ \cite{gasperini,obukhov}. The
constant $\al$ is any number.

The Dirac action $S_\psi$ is given by
\beq
S_\psi = \frac{i}{2}\int d^4x\sqrt{-g}\left\{\bar{\psi}\ga^\mu\tilde{\nabla}_\mu\,\psi -
\tilde{\nabla}_\mu\,\bar{\psi}\ga^\mu\psi + 2im\bar{\psi}\psi \right\}\,,
\eeq
where the operator $\tilde{\nabla}_\mu$ on spinors is defined using the spinor connection
$$
\tilde{\omega}^{ab}\mbox{}_\mu = \frac{1}{2}\left(
e^{b\al}\partial_\mu e^a\mbox{}_\al
- e^{a\al}\partial_\mu e^b\mbox{}_\al \right)
- \frac{1}{2} \tilde{\Gamma}^\rho\mbox{}_{\al\mu}
\left( e^{b\al} e^a\mbox{}_\rho - e^{a\al} e^b\mbox{}_\rho
\right)\, .
$$
In the above expression, the objects $e^a\mbox{}_\al$ are the
vierbeins such that $\ga_\al = e^a\mbox{}_\al\,\ga_a$, $e^a\mbox{}_\al\, e^{b\al} = \eta^{ab}$
and $e^a\mbox{}_\al\, e_{a\be} = g_{\al\be}$. Separating the Riemaniann part from the torsion part,
we can express the Dirac action after some algebraic manipulation as follows \cite{torsi}:
\beq
S_\psi = \frac{i}{2}\int d^4x\sqrt{-g}\left\{\bar{\psi}\ga^\mu\nabla_\mu\,\psi -
\nabla_\mu\,\bar{\psi}\ga^\mu\psi + 2im\bar{\psi}\psi\right\} +
\int d^4x\sqrt{-g}\frac{1}{8}S_\mu\,J^\mu\,,
\eeq
where $J^\mu$ is the axial current $\bar{\psi}\ga^5\ga^\mu\psi$ and $S_\mu$ is the
pseudo-trace of torsion which can be seen in decomposition of torsion in its
irreducible parts (along with trace $T_\al := T^\mu\mbox{}_{\al\mu}$ and tensorial part
$q^\mu\mbox{}_{\al\be}$),
\beq
T^\mu\mbox{}_{\al\be} = \frac{1}{3}\left( \de^\mu_\be T_\al - \de^\mu_\al T_\be\right)
-\frac{1}{6} \ep^\mu\mbox{}_{\al\be\si}S^\si + q^\mu\mbox{}_{\al\be}\,. \label{decomp}
\eeq
The pseudo-trace and tensor $q^\mu\mbox{}_{\al\be}$ satisfy the identities
$$
\ep^{\mu\al\be\si}\,T_{\mu\al\be} = S^\si \,,\;\;\;\;\;
q^\mu\mbox{}_{\al\be} = -q^\mu\mbox{}_{\be\al}\;\;\; {\rm and} \;\;\;
\ep^{\mu\al\be\si}\,q_{\mu\al\be} = 0\,.
$$

We have so far the total action with all terms defined such that
in order to write separatelly the torsion independent part and
the torsion dependent part, we should take into account
\beq
\tilde{R}^\tau\mbox{}_{\si\rho\la} = R^\tau\mbox{}_{\si\rho\la} +
K^\tau\mbox{}_{\phi\rho}K^\phi\mbox{}_{\si\la} -
K^\tau\mbox{}_{\phi\la}K^\phi\mbox{}_{\si\rho} + \; {\rm total}
\;\; {\rm derivatives}\,.
\eeq

Now we are able to vary the total action in relation to
$K^\mu\mbox{}_{\al\be}$ to obtain the equation for torsion.
Then, we shall substitute torsion by its sources back into the
action.

\section{Equation for torsion and equivalent action}

Let us write some useful computations in varying the action. After straightforward
calculation, one achieves
\beq
\frac{\de}{\de K^\mu\mbox{}_{\al\be}}\int d^4x\sqrt{-g}\,(-\tilde{R}) =
\de^\be_\mu T^\al - g^{\al\be} T_\mu - K_\mu\mbox{}^{\be\al} +
K^{\al\be}\mbox{}_\mu\,,
\eeq
\beq
\frac{\de}{\de K^\mu\mbox{}_{\al\be}}\int d^4x\sqrt{-g}\,
\tilde{R}^{\tau\si\rho\la}\ep_{\tau\si\rho\la} =
2\ep_\mu\mbox{}^{\si\be\la}K^\al\mbox{}_{\si\la} +
2\ep^{\al\si\rho\be}K_{\mu\si\rho}\,,
\eeq
and
\beq
\frac{\de}{\de K^\mu\mbox{}_{\al\be}}\int d^4x\sqrt{-g}\,
S_\rho J^\rho = 2\ep_\mu\mbox{}^{\al\be\si}J_\si\,.
\eeq
Using these results, we obtain the equation for torsion, $\de S/\de K^{\al\be}\mbox{}_\mu = 0$,
in the form
\beq
& & \frac{1}{\ka}\left\{
\de^\be_\mu T^\al - g^{\al\be} T_\mu + T^\be\mbox{}_\mu\mbox{}^\al + \frac{1}{\be}\left(
2\ep_\mu\mbox{}^{\si\be\la}K^\al\mbox{}_{\si\la} +
2\ep^{\al\si\rho\be}K_{\mu\si\rho}\right)\right\} \nonumber \\
& = & -\al\Si_\mu\mbox{}^\al U^\be -
2\eta\ep_\mu\mbox{}^{\al\be\si}J_\si\,.
\label{torsion}
\eeq
Contracting $\be$ with $\mu$, one gets
\beq
T^\al = \frac{1}{4\be} S^\al\,, \label{t}
\eeq
which is also useful to achieve the following result after multiplying (\ref{torsion})
by $\ep_{\be\mu\al\tau}$:
\beq
S_\tau = -\al\ka\theta\ep_{\be\mu\al\tau}\Si^{\mu\al}U^\be + 12\eta\ka\theta J_\tau\,,
\label{s}
\eeq
where
$$
\theta = \frac{\be^2}{\be^2 + 1}\,.
$$

We can use the above equations in (\ref{torsion}) and after some algebraic manipulations
we get finally
\beq
T^\mu\mbox{}_{\al\be} & = & \frac{\al\ka\theta}{4\be}\left(\de^\mu_\be \ep_{\al\rho\la\tau}\Si^{\rho\la}U^\tau -
\de^\mu_\al \ep_{\be\rho\la\tau}\Si^{\rho\la}U^\tau - 2\ep_{\al\be\rho\la}\Si^{\rho\la}U^\mu\right) + \nonumber \\
& + & \frac{\eta\ka\theta}{\be}\left( \de^\mu_\be J_\al - \de^\mu_\al J_\be \right) +
\al\ka\theta \Si_{\be\al}U^\mu + 2\eta\ka\theta \,\ep^{\mu\si}\mbox{}_{\be\al}J_\si\,.
\label{torsion2}
\eeq

Now we are in a position to substitute all torsion terms in the action (\ref{action})
by the sources $\Si^{\rho\la}$ and $J_\nu$ through the equation (\ref{torsion2}). This
task is quite tedious, so the best way to do this is to rewrite the action in terms of
Riemannian structures and the irreducible components of torsion, $T_\mu$, $S_\mu$ and
$q^\mu\mbox{}_{\al\be}$, and then not to substitute equation (\ref{torsion2}) directly,
but the corresponding equations for components of torsion, which can be derived by taking
into account equations (\ref{decomp}), (\ref{t}), (\ref{s}) and (\ref{torsion2}). For both
$T_\mu$ and $S_\mu$, we already have the convenient expressions (equations (\ref{t}), (\ref{s})). For
the remaining component, $q^\mu\mbox{}_{\al\be}$, we obtain
\beq
q^\mu\mbox{}_{\al\be} & = & \frac{\al\ka\theta}{6\be}\left(\de^\mu_\be \ep_{\al\rho\la\tau}\Si^{\rho\la}U^\tau -
\de^\mu_\al \ep_{\be\rho\la\tau}\Si^{\rho\la}U^\tau - 3\ep_{\al\be\rho\la}\Si^{\rho\la}U^\mu\right)  \nonumber \\
& - & \frac{\al\ka\theta}{3} \left( \Si^\mu\mbox{}_\be U_\al - \Si^\mu\mbox{}_\al U_\be
+ 2\Si_{\al\be}U^\mu \right)\,.
\label{q}
\eeq

Actually it would be much more simple to write the action, in the very beginning,
in terms of the irreducible components, and then take the variation of the action
in relation to $\de T_\mu$, $\de S_\mu$ and $\de q^\mu\mbox{}_{\al\be}$
separatelly. Nevertheless, this procedure is not equivalent to what we have done
because the components $T_\mu$, $S_\mu$ and $q^\mu\mbox{}_{\al\be}$ are not
independent each other\footnote{In this alternative and simpler way to derive the
equations for torsion, the equation for $q^\mu\mbox{}_{\al\be}$, analogous to
(\ref{q}), would be such that $\ep^{\mu\al\be\rho}q_{\mu\al\be} \neq 0$, what
violates the defining property of $q^\mu\mbox{}_{\al\be}$.}.

Thus after straighforward calculations, with the help of the useful
equations
\beq
-\tilde{R} = -R + \frac{2}{3}T_\mu T^\mu - \frac{1}{24}S_\mu S^\mu
-\frac{1}{2}q_{\mu\nu\rho}q^{\mu\nu\rho}\,
\eeq
and
\beq
\frac{1}{2\be}\ep^{\mu\nu\rho\la}\tilde{R}_{\mu\nu\rho\la} =
-\frac{1}{3\be}T_\mu S^\mu +
\frac{1}{4\be}\ep^{\mu\nu\al\be}q_{\si\mu\nu} q^\si\mbox{}_{\al\be}\,,
\eeq
one finally can write the full action eliminating all torsion terms as
\beq
S & = & S_{EH} + S_{PF} + S_{{\rm Dirac}} + \\
& + & \ka\theta\int d^4x\sqrt{-g}\left\{
\frac{\al^2}{4}\Si_{\al\be}\Si^{\al\be} +
\al\eta\,\ep_{\mu\nu\al\be}J^\mu\Si^{\nu\al} U^\be +
6\eta^2J_\mu J^\mu\right\}\,, \nonumber
\label{final}
\eeq
where $S_{PF}$ is the perfect fluid action and
$$
S_{EH} = -\frac{1}{\ka}\int d^4x\sqrt{-g}\,R\,,
$$
$$
S_{{\rm Dirac}} = \frac{i}{2}\int d^4x\sqrt{-g}\left\{\bar{\psi}\ga^\mu\nabla_\mu\,\psi -
\nabla_\mu\,\bar{\psi}\ga^\mu\psi + 2im\bar{\psi}\psi\right\}\,.
$$

In the above result, the choice $\eta = 1/8$ (minimal coupling)
implies the coefficient $3\pi G\theta/2$ for the 4-fermion term
$J_\mu J^\mu$. In Ref. \cite{perezrovelli} this coefficient is
the same\footnote{Except by a non-essential sign coming from different
notations.} (see also \cite{BPFSS}).

\section{Conclusions}

We have considered the Holst action with fermions and a Weyssenhoff
fluid in the spacetime with torsion. Torsion is responsible for
the existence of non-trivial effects of the term
$\ep^{\mu\nu\al\be}\,\tilde{R}_{\mu\nu\al\be}$ through its
coefficient $1/2\be$. By varying the action in respect to torsion
(or contortion), one can write down the dynamical equation for torsion,
which actually is not dynamical because it just relates algebraically
torsion with the sources (fermionic current and intrinsic spin), such
that we found the equivalent action in the form of Einstein-Hilbert
gravity in Riemannian spacetime with fermions and a perfect fluid,
together to the additional term
$$
\ka\theta\int d^4x\sqrt{-g}\left\{
\frac{\al^2}{4}\Si_{\al\be}\Si^{\al\be} +
\al\eta\,\ep_{\mu\nu\al\be}J^\mu\Si^{\nu\al} U^\be +
6\eta^2J_\mu J^\mu\right\}\,.
$$

The fluids are macroscopic entities, such the intrinsic spin of the
fluid is conveniently thought as a quantity subject to statistical
procedures. In the literature, the relevant quantity is defined as
\beq
2\si^2 = <\Si_{\mu\nu}\Si^{\mu\nu} >\,,
\eeq
in such a way that $<\Si_{\mu\nu}>$ should vanish because
os its random character. Notice that the current $J_\mu$
is different because it can be treated as a background external
preferred direction, which is not random. In what sense
a coupling between an external quantity, non-random, and
a random quantity, as $\ep_{\mu\nu\al\be}J^\mu\Si^{\nu\al} U^\be$,
can keep $\Si^{\nu\al}$ random? The answering of this and similar
questions we postpone for another work. Let us note that the $J_\mu$
in the form $(0,J_x,J_y,J_z)$ can make
$\ep_{\mu\nu\al\be}J^\mu\Si^{\nu\al} U^\be$ non-trivial and thus
inducing some anisotropy in the cosmological context, opening the
possibility of establishing observational limits to
$\al\eta\theta$.

\subsection*{Acknowledgments}

G.B.P. is grateful to CNPq and FAPEMIG for partial support. C.A.S.
is grateful to CAPES for PhD program. We acknowledge the comments and
suggestions from the audience of our talk in the seminar session of
our research group.



\end{document}